\documentclass[11pt]{iopart}

\usepackage[british]{babel}
\usepackage{color}
\usepackage{mathrsfs}
\usepackage{setstack}
\usepackage{iopams}
\usepackage[T1]{fontenc}
\usepackage{graphicx}
\usepackage{color}
\usepackage[sort&compress,numbers]{natbib}

\usepackage{amssymb}
\usepackage{nicefrac}

\newcommand{\figref}[1]{Fig.~\ref{#1}}

\newcommand{\eqref}[1]{(\ref{#1})}

\newcommand{\bd}[1]{\textcolor{blue}{#1}}
\renewcommand{\bd}[1]{\textcolor{black}{#1}}

\newcommand{\tmax}{t_{\mathrm{max}}}

\newcommand{\pd}[3]{\frac{ \partial^{ #1 } #2 }{ \partial #3^{ #1 } }}

\begin{document}

\title{Multimodal stationary states in symmetric single-well potentials driven by Cauchy noise}
\author{Karol Capa{\l}a, Bart{\l}omiej Dybiec}
\address{Marian Smoluchowski Institute of Physics,
and Mark Kac Center for Complex Systems Research, Jagiellonian University, ul. St. {\L}ojasiewicza 11, 30--348 Krak\'ow, Poland}
\ead{karol@th.if.uj.edu.pl, bartek@th.if.uj.edu.pl}

\begin{abstract}
Stationary states for a particle moving in a single-well, steeper than parabolic, potential driven by L\'evy noise can be bi-modal.
Here, we explore in details conditions that are required in order to induce multimodal stationary states having more than two modal values.
Phenomenological arguments determining necessary conditions for emergence of stationary states of higher multimodality are provided.
Basing on these arguments, appropriate symmetric single-well potentials are constructed.
Finally, using numerical methods it is verified that stationary states have anticipated multimodality.
\end{abstract}


%pacs{
% 05.40.Fb, % Random walks and Levy flights
% 05.10.Gg, % Stochastic analysis methods (Fokker--Planck, Langevin, etc.)
% 02.50.-r, % Probability theory, stochastic processes, and statistics
% 02.50.Ey, % Stochastic processes
% }

\maketitle

%%%%%%%%%%%%%%%%%%%%%%%%%%%%%%%%%%%%%%%%%%%%%%%%%%%%%
%%%%%%%%%%%%%%%%%%%%%%%%%%%%%%%%%%%%%%%%%%%%%%%%%%%%%
\section{Introduction\label{sec:introduction}}

A particle immersed into a liquid constantly interacts with other particles of its environment.
These collision results in the irregular, observable motion of a test particle, which is called the Brownian motion.
The theory of the Brownian motion  had been rigorously and independently developed by Einstein \cite{Einstein1905} and Smoluchowski \cite{Smoluchowski1906} in series of papers which made fundamental contributions to kinetic theory of matter, theory of fluctuations and nonequilibrium statistical mechanics.

The Brownian motion is an example of a continuous-space and a continuous-time Markov process with independent increments.
Brownian motion can be described by the simplest form of the overdamped Langevin equation
\begin{equation}
\dot{x}=\xi(t),    
\label{eq:lw-wiener}
\end{equation}
where $\xi(t)$ is the Gaussian white noise.
In the theory of stochastic processes, the Brownian motion is the Wiener process.
In Eq.~(\ref{eq:lw-wiener}), $\xi(t)$ is the so called noise which is used as efficient way of approximation of not fully known interactions of a test particle with its environment. 
Increasing number of observations suggest that fluctuations in real systems do not need to follow the Gaussian law.
The natural generalization of the Gaussian density is provided by the $\alpha$-stable density \cite{janicki1994,samorodnitsky1994} which lead to the so called L\'evy flights.

L\'evy flights correspond to summation of uncorrelated random steps drawn from a heavy tailed density.
Typically it is assumed that tails are of the power-law type.
Therefore, L\'evy flights extends Brownian motion for which the step increments are Gaussian. A simple scaling argument shows that, unlike Gaussian Wiener process, L\'evy flights are characterized by infinite variance, so that the width of the diffusive ``packet'' must be understood in terms of some fractional moments or the interquartile distance \cite{dybiec2008d}. 
The infinite variance of free L\'evy flights is responsible for peculiar properties of systems driven by L\'evy noise, see  \cite{dubkov2008,metzler2007} and below.
\bd{Therefore, the scenario of L\'evy flights should be contrasted with the complementary model of L\'evy walks \cite{zaburdaev2015levy}, which assures finite and constant propagation velocity.}

\bd{Theory of L\'evy flights has been developed in a series of papers including among others \cite{metzler2000,barkai2001,anh2003,brockmann2002,chechkin2006,jespersen1999,yanovsky2000,schertzer2001}.}
L\'evy flights  have been studied in various contexts \cite{klages2008} with applications ranging from economy and finance \cite{bouchaud1990} to superdiffusion of micellar systems \cite{bouchaud1991}, studies of turbulence \cite{shlesinger1986b}, description of photons in hot atomic vapors \cite{mercadier2009levyflights} and laser cooling \cite{cohen1990,barkai2014}.
These studies included not only experimental \cite{solomon1993} but also various  theoretical aspects  \cite{eliazar2003,sokolov2004b,garbaczewski2009,garbaczewski2010} of L\'evy flights.

A test particle might be  not only perturbed by the noise but also it can be driven by the deterministic force, e.g. $f(x)=-V'(x)$, which is to be added to the right hand side of Eq.~(\ref{eq:lw-wiener}).
The overdamped Langevin equation
\begin{equation}
\dot{x}(t) = - V'(x) + \xi(t)
    \label{eq:EoM}
\end{equation}
is a one of fundamental equation in theory of stochastic systems.
In Eq.~(\ref{eq:EoM}), $-V'(x)$ stands  for the deterministic force, while $\xi(t)$ denotes, as in Eq.~(\ref{eq:lw-wiener}), the stochastic term.
In the limit of vanishing noise strength the motion of an overdamped particle becomes deterministic and especially simple.
The deterministic force drives a particle along the potential slope, e.g. to a minimum of the potential which is stable.
Presence of noise introduces randomization of trajectories.
Moreover, it changes stability of minima of the potential.
Combined action of a Gaussian white noise and deterministic forces produces stationary states in potential wells which are of the Boltzman Gibbs type.
Replacement of the Gaussian  driving with L\'evy noise significantly alters conditions for existence, properties and shape of stationary states \cite{chechkin2002,chechkin2003,chechkin2004,chechkin2006,chechkin2008introduction}.
\bd{Eq.~(\ref{eq:EoM}) provides foundations of the Kramers rate theory \cite{kramers1940,hanggi1990}, which can be also generalized to the non-equilibrium, L\'evy flight regime \cite{chechkin2005,dybiec2007}.}
Furthermore, Eq.~(\ref{eq:EoM}) can be extended for time dependent forces resulting in plenitude of noise-induced effects like stochastic resonance \cite{gammaitoni2009}, resonant activation \cite{doering1992} and ratcheting effect \cite{reimann2002}.

In this paper we are analyzing the problem of stationary states in stochastic systems described by an overdamped Langeving equation.
The problem we aim to address is whether action of nonequilibrium noises of $\alpha$-stable type can induce multimodal stationary states in symmetric single-well potentials. 
Despite the fact that the conclusive answer to this problem is known for a long time, it is still unknown whether it is possible to observe multimodal states characterized by more than two modal values.
Within the current manuscript we are filling this gap.
In the next section (Sec~\ref{sec:model}) a model under studies with required theory is presented. Section Results~(Sec.~\ref{sec:results}) provides obtained results proving that for a properly tailored symmetric single-well potentials one can produce tri- and higher modal stationary states.
The manuscript is closed with Summary and Conclusions (Sec.~\ref{sec:summary}).

%%%%%%%%%%%%%%%%%%%%%%%%%%%%%%%%%%%%%%%%%%%%%%%%%%%%%
%%%%%%%%%%%%%%%%%%%%%%%%%%%%%%%%%%%%%%%%%%%%%%%%%%%%%
\section{Model\label{sec:model}}

We are studying the system described by Eq.~(\ref{eq:EoM}) in which the Gaussian white noise is replaced by a more general noise of the $\alpha$-stable type \cite{peszat2007stochastic,chechkin2008introduction,metzler2007}.
From the whole class of $\alpha$-stable noises we concentrate on symmetric ones \cite{samorodnitsky1994,janicki1994,dubkov2008}.
$\alpha$-stable white noise is a formal time derivative of the $\alpha$-stable process, whose increments are distributed according to an $\alpha$-stable density, which is uni-modal, heavy-tailed probability density.
The characteristic function of the symmetric $\alpha$-stable variables \cite{samorodnitsky1994,janicki1994} is given by
\begin{equation}
\varphi(k)=e^{i\sigma^{\alpha}|k|^{\alpha}},
    \label{eq:levycf}
\end{equation}
where $\alpha$ ($0<\alpha \leqslant 2$) is the stability index determining the exponent characterizing power-law decay of $\alpha$-stable densities, which for $\alpha<2$ is of $ |x|^{-(\alpha+1)}$ type. The scale parameter $\sigma$ controls the distribution width.
For $\alpha<2$, the variance of an $\alpha$-stable density is infinite, thus the distribution width needs to be understood as the interquantile width.
For $\alpha=2$, the characteristic function (\ref{eq:levycf}) reduces to the characteristic function of  the normal (Gaussian) density. The case of $\alpha=1$ corresponds to the Cauchy distribution.
Increments of an $\alpha$-stable process are distributed according to the $\alpha$-stable density with the characteristic function $\varphi(k)=\exp({i\Delta t\sigma^{\alpha}|k|^{\alpha}})$.
Therefore, the Langevin equation (\ref{eq:EoM}) can be approximated by \cite{janicki1994,janicki1996}
\begin{equation}
    x(t+\Delta t) = x(t) - V'(x) \Delta t + \xi_t \Delta t^{1/\alpha},
\end{equation}
where $\xi_t$ represents a sequence of independent identically distributed random variables \cite{chambers1976,weron1995,weron1996} following the $\alpha$-stable density, see Eq.~(\ref{eq:levycf}).

 Complementary to the Langevin equation, the evolution of the probability density is described by the fractional Smoluchowski-Fokker-Planck equation \cite{samorodnitsky1994,podlubny1998,yanovsky2000}
\begin{equation}
 \pd{}{p(x,t)}{t} = -\pd{}{}{x} V'(x,t)p(x,t) + \sigma^{\alpha} \pd{\alpha}{p(x,t)}{|x|},
 \end{equation}
 where the fractional Riesz-Weil derivative \cite{podlubny1998,samko1993} is defined via the Fourier transform $
 \mathcal{F}_k\left( \frac{\partial^\alpha f(x)}{\partial |x|^\alpha} \right)=-|k|^\alpha \mathcal{F}_k\left(f(x)\right).
$

For systems driven by the Gaussian white noise, i.e. the $\alpha$-stable noise with $\alpha=2$, stationary states exist for any potential $V(x)$ having the property $V(x) \to \infty$ as $|x|\to~\infty$. 
Moreover, stationary states are of the Boltzmann-Gibbs type, i.e. $\ln p(x) \propto - V(x)$.
Systems driven by an $\alpha$-stable noise display very different properties than their Gaussian white noise-driven counterparts.
First of all, for a single-well potential of $|x|^n$ type, the exponent $n$ characterizing the steepness of the potential needs to be large enough in order to produce stationary states \cite{dybiec2010d}. More precisely, the following relation needs to be satisfied
\begin{equation}
    n > 2 -\alpha.
\end{equation}
Furthermore, if a stationary state exists it is not of the Boltzmann-Gibbs type \cite{eliazar2003}.

\begin{figure}[h!]
\centering
\includegraphics[width=0.7\textwidth]{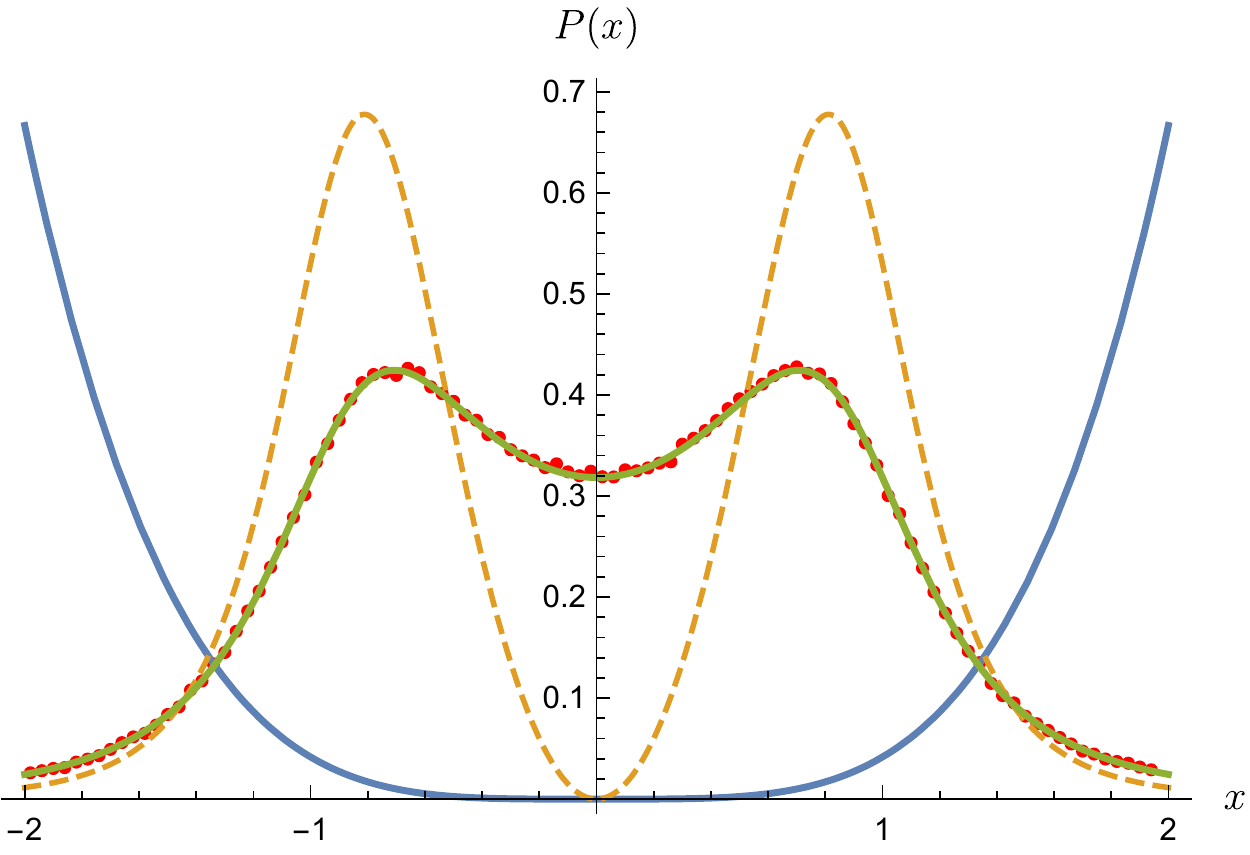}
\caption{Stationary state for the quartic potential $V(x)=x^4/4$ subject to action of the $\alpha$-stable driving with $\alpha$=1 (Cauchy noise) and $\sigma=1$ (dots), 
quartic potential (blue solid line), exact solution \eqref{eq:n4} (green solid line) and the potential curvature (orange dashed line). 
Simulation parameters: $N=10^6$, $\Delta t =10^{-3}$ and $\tmax=100$.}
\label{fig:n4}
\end{figure}

In analogy with systems driven by Gaussian white noise, for $n=2$ stationary states for systems driven by $\alpha$-stable noise reproduce the noise distribution. 
Therefore, stationary states are given by the rescaled $\alpha$-stable density with the same stability index $\alpha$ like the driving noise \cite{chechkin2003}.
In addition to $n=2$, the formulas for the stationary state is known for $V(x)=x^4/4$ driven by the Cauchy noise, i.e. the $\alpha$-stable noise with $\alpha=1$. The appropriate formula \cite{chechkin2002,chechkin2003,chechkin2004} for the stationary density reads
\begin{equation}
    p_{\alpha=1}(x)=\frac{1}{\pi\sigma^{\nicefrac{1}{3}}\left[(x/\sigma^{\nicefrac{1}{3}})^4-(x/\sigma^{\nicefrac{1}{3}})^2+1\right]}.
    \label{eq:n4}
\end{equation}
The probability density (\ref{eq:n4}) is a bi-modal density. Bi-modality is a general property of stationary states in steeper than parabolic potentials subject to the action of L\'evy noises \cite{chechkin2002,chechkin2003}.
Consequently, the stationary state produced in a symmetric single-well potential does not reproduce the symmetry of the potential, i.e. it is not uni-modal, see Fig.~\ref{fig:n4}.
The transition between uni-modal and bi-modal stationary states takes place at $n=2$.
For $n>2$ stationary states are bi-modal with the minimum at the origin, while for $2-\alpha<n<2$ they are uni-modal with the maximum at the origin.
In double-well potentials, stationary states are bi-modal \cite{dybiec2007d,sliusarenko2012}
Finally, for an infinite rectangular potential well the stationary state is
\begin{equation}
    p(x)=\frac{\Gamma(\alpha)(2L)^{1-\alpha}(L^2-x^2)^{\alpha/2-1}}{\Gamma^2(\alpha)},
    \label{eq:rw}
\end{equation}
see \cite{denisov2008}, where $\Gamma(\dots)$ is the Euler-Gamma function.
In the special case of  $\alpha=1$, the stationary state, see Eq.~(\ref{eq:rw}), is given by the arcsin distribution.

The infinite rectangular potential well with impenetrable boundaries located at $x=\pm L$ can be obtained as $n\to\infty$ limit of the symmetric single-well potential
\begin{equation}
    V_n(x)=\frac{(x/L)^{2n}}{2n},
    \label{eq:vn}
\end{equation}
for which stationary states for $\alpha=1$ are also known  \cite{dubkov2007}.
Detailed examination of such a transition allow one to see how stationary states in systems driven by $\alpha$-stable noise emerges \cite{kharcheva2016}.
Additionally, Eq.~(\ref{eq:n4}) provides an example that demonstrates the influence of the scale parameter on the shape of stationary states.
From Eq.~(\ref{eq:n4}), one can conclude that with the increasing $\sigma$ maxima of probability density~(\ref{eq:n4}) shift toward larger absolute values of arguments.
This displacement is related to the fact that for the large $\sigma$ more probability mass is shifted to the tails of the jump length distribution, thus making central part of the distribution less prominent.
Analogously, the increase in the exponent characterizing steepness of the potential $n$, see Eq.~(\ref{eq:vn}), moves maxima of the stationary states towards $x=\pm L$. 
Finally, in the limit of $n=\infty$ maxima are located exactly at boundaries, i.e. $\pm L$, see Eq.~(\ref{eq:rw}).

In short, the stationary state  is determined by the interplay between random and deterministic forces. The deterministic force is defined by a static potential $V(x)$, while a stochastic force arises due to noise. 
Since the studied system is overdamped, see Eq.~(\ref{eq:EoM}), the deterministic force is responsible for sliding down of a particle to the potential minimum.
The random force resulting from the noise $\xi(t)$ is the only force which can move a particle away from the minimum of the potential.
Therefore, the competition between random excursion and deterministic sliding  defines the shape of a stationary state.
This mechanism is related to the decomposition of $\alpha$-stable noises \cite{imkeller2006,imkeller2006b,samorodnitsky2007}.

The above mentioned description presents a general mechanism responsible for the shape of stationary states. This mechanism is of very general type and as a such do not provide simple estimates for positions of maxima of stationary states.
Therefore, we extend a phenomenological arguments \cite{chechkin2002,chechkin2004} which could provide more information about stationary states, e.g. about positions of modal values.

As a test bench we use the quartic $x^4/4$ potential well perturbed by the Cauchy noise with $\sigma=1$, for which the stationary state is given by Eq.~(\ref{eq:n4}).
Fig.~\ref{fig:n4} presents results of computer simulations (dots), the quartic potential (blue solid line) and the potential curvature (orange dashed line).
For a plane curve given by  $V(x)$, the curvature is 
\begin{equation}
\kappa(x)=\frac{V''(x)}{[1+V'(x)^2]^{3/2}}.
    \label{eq:curvature}
\end{equation}
In order to increase clarity of Fig.~\ref{fig:n4}, the curvature $\kappa(x)$ is divided by a constant.
From Fig.~\ref{fig:n4} it is clearly visible that maxima of stationary state are located closely to the maxima of the curvature of $V(x)$, see \cite{chechkin2002,chechkin2004}.
For further reference, let us call $\bar{x}_i$ $i$-th value of $x$ for which curvature $ \kappa(x)$ has its local maximum. 
The significant likelihood of concentrating of probability mass near maxima of curvature comes from the fact that maximum of curvature describes a point where a transition from dominance of almost vertical to flat behaviour of the potential takes place. 
This is especially well visible for the infinite rectangular potential well because at the point of maximal curvature the potential changes from the horizontal to the vertical (reflecting boundary).
Potential slope is directly connected with the change of a particle position, see Eq.~\eqref{eq:EoM}, while the curvature describes how rapidly movement of the particle changes at a small distance. 
Therefore, a maximum of the curvature establish a point where a change of a particle position is the most hampered.  
Please notice, that the conjecture  associating maxima of the potential curvature with modal values of stationary states \cite{chechkin2004} confirm also uni-modal --- bi-modal transition at $n=2$ for single-well potentials of $|x|^n/n$ type.

Due to peculiar properties of systems driven by $\alpha$-stable noises one can inquire about possibility of producing multimodal stationary states in symmetric single-well potentials.
In the following section we show that for properly tailored symmetric single-well potentials it is possible to produce stationary states having more than two modal values.

%%%%%%%%%%%%%%%%%%%%%%%%%%%%%%%%%%%%%%%%%%%%%%%%%%%%%
%%%%%%%%%%%%%%%%%%%%%%%%%%%%%%%%%%%%%%%%%%%%%%%%%%%%%
\section{Results\label{sec:results}}

In this section we show numerically that for special types of symmetric single-well potentials it is possible to obtain multimodal stationary states characterized by more than two modal values.
In particular, we demonstrate sample symmetric, differentiable, single-well potentials (Sec.~\ref{sec:dif}) resulting in three-modal, four-modal  and five-modal stationary states.
Moreover, we show ``glued'' symmetric single-well potentials (Sec.~\ref{sec:non-dif}) which are also able to produce multimodal stationary states.
All considered potentials are symmetric single-well, i.e. the only one minimum of the potential is located at $x=0$. 
Importantly, for $x>0$ ($x<0$) potentials are non-decaying (non-decreasing) functions of $x$ 
with the non-monotonous dependence of the curvature characterized by several maxima.

\bd{
Sample potentials used in further numerical studies are pre-selected by the curvature condition and then numerically fine-tuned.
The requirement of several maxima of the potential curvature is necessary to obtain a multimodal stationary state.
Unfortunately, this condition solely could be not sufficient.
For instance, maxima of the potential curvature need to be adequately separated.
Otherwise, maxima of a stationary state can interfere resulting in smaller number of peaks in the stationary state than the number of maxima in the potential curvature. 
Therefore, the condition on the number of maxima of the potential curvature needs to be accompanied by additional numerical tests.
Exemplary values of the potential parameters, used in further studies, represent sample values of coefficients resulting in pronounced stationary states having required number of modal values.
}

%%%%%%%%%%%%%%%%%%%%%%%%%%%%%%%%%%%%%%%%%%%%%%%%%%%%%
%%%%%%%%%%%%%%%%%%%%%%%%%%%%%%%%%%%%%%%%%%%%%%%%%%%%%
\subsection{Continuous differentiable potentials\label{sec:dif}}

%%%%%%%%%%%%%%%%%%%%%%%%%%%%%%%%%%%%%%%%%%%%%%%%%%%%%
\subsubsection*{Three mods\label{sec:3mods}}

Let us start with the overdamped Langevin equation (\ref{eq:EoM}) with the symmetric single-well potential
\begin{equation}
    V(x)=x^2-ax^4+x^6.
    \label{eq:V1}
\end{equation}
The above potential has one minimum located at $x=0$ and three points in which the curvature has its local maxima.
We use numerical methods in order to find stationary states for the system described by Eq.~(\ref{eq:EoM}).

%%%%%%%%%%%%%%%%%%%%%%%%%%%%
%
% fig
%
\begin{figure}[h!]
\centering
\includegraphics[width=0.7\textwidth]{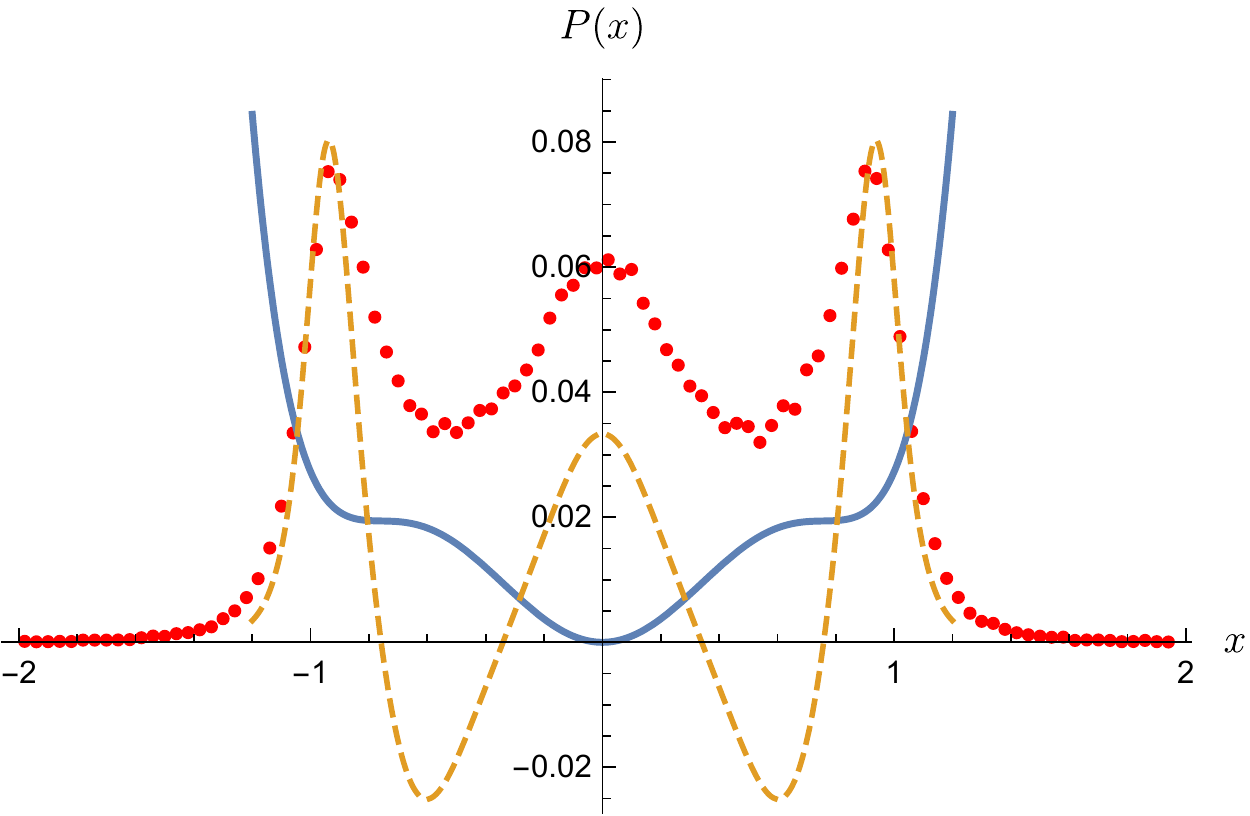}
\caption{The stationary state for the potential given by Eq.~\eqref{eq:V1} with $a=19/11$ subject to the $\alpha$-stable driving with $\alpha$=1 (Cauchy noise) and $\sigma=1$  (dots), the potential profile (blue solid line) and the potential curvature (orange dashed line). Simulation parameters: $N=10^6$, $\Delta t =10^{-3}$, $\tmax=100$.}
\label{fig:3mods}
\end{figure}

Numerical results for the potential \eqref{eq:V1} with $a=19/11$ together with the potential profile are shown in \figref{fig:3mods}. 
As it is visible in Fig.~\ref{fig:3mods}, the stationary state has three well visible peaks. One of them is located at the minimum of the potential at $x=0$ which is also the local maximum of the potential curvature. 
For small value of $x$ the potential (\ref{eq:V1}) can be approximated by the parabolic part.
Therefore, dynamics of a particle at $x \approx 0$ is of the same type like the motion in the parabolic potential, which results in the emergence of the single peak at $x=0$.
Two other peaks appear near to two remaining maxima of curvature of the potential due to action of the deterministic force produced by outer ($|x|>1$) parts of the potential.

\bd{
The parameter $a$ in Eq.~\eqref{eq:V1} is adjusted to assure that the potential $V(x)$ is still of a single-well type and the stationary state is tri-modal.
For $a=19/11$ the potential~\eqref{eq:V1} is close of having three minima, since for $a> \sqrt{3}$ the potential has three minima. In the case of $a>\sqrt{3}$, the stationary state is tri-modal.
Consequently, $a= \sqrt{3}$ gives the upper bound of the domain of the $a$ parameter. For $a<0$, the stationary state is unimodal. Therefore, we had to consider $0 < a <  \sqrt{3}$. 
Finally, we have performed additional simulations in order to find the critical value of the $a$ parameter for a transition between tri-modal and unimodal stationary state. From our simulations, we see that the critical value is located between $ 1.17 < a_c \leq 1.20$. Unfortunately, due to inherent uncertainties of our methodology, it is very hard to provide a better estimate. 
}

%%%%%%%%%%%%%%%%%%%%%%%%%%%%
%
% fig
%
\begin{figure}[h!]
\centering
\includegraphics[width=0.7\textwidth]{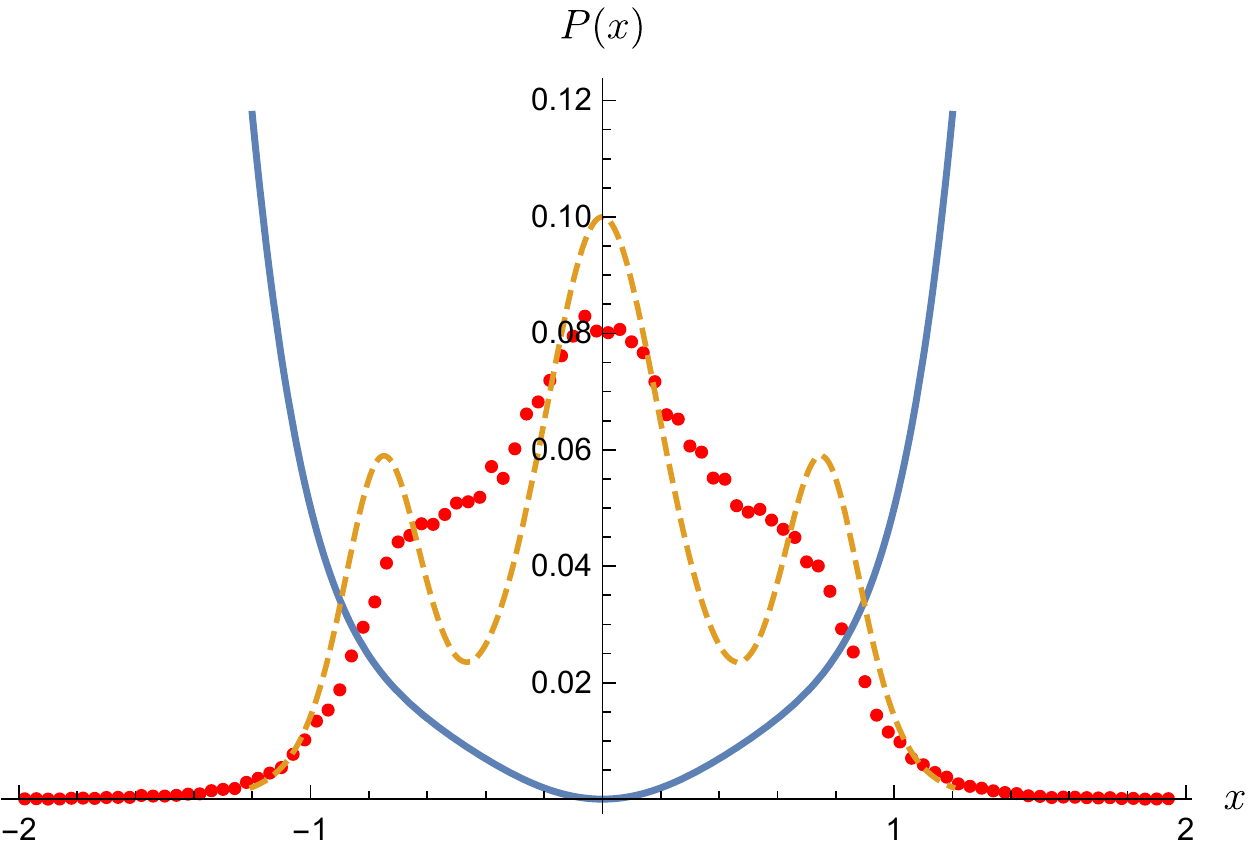}
\caption{The same as in \figref{fig:3mods} for $a=1$.}
\label{fig:1mod}
\end{figure}

Different situations is observed for the potential given by Eq.~\eqref{eq:V1} with $a=1$, which is smaller than $a_c$, see \figref{fig:1mod}. 
The potential still has one minimum at $x=0$.
The curvature has three maxima: $\bar{x}_0$ and $\bar{x}_\pm$.
The maximum $\bar{x}_0$ located at $x=0$ is dominating and remaining points of maximal curvature $\bar{x}_{\pm 1}$ are closer to each other than for $a=19/11$, see Fig.~\ref{fig:3mods}.
Relative changes in the curvature are also smaller than for $a=19/11$. 
In Fig.~\ref{fig:1mod}, there is only one maximum of the stationary density located at the origin because of different curvature profile than in Fig.~\ref{fig:3mods}.
In other words, maxima of curvature are not distanced (separated) enough to induce tri-modal stationary state.
Nevertheless, the influence of curvature is still visible.
More precisely, for $x<0$, in the range where curvature decays from its local maximum $\bar{x}_-$ to its global minimum the stationary density increases slower than in the areas of growing curvature.
The very similar effect is observed for $x>0$, where the decay of the stationary density is slower in the interval where the curvature grows.

%%%%%%%%%%%%%%%%%%%%%%%%%%%%%%%%%%%%%%%%%%%%%%%%%%%%%%%%%%%%%%%%%
\subsubsection*{Four mods\label{sec:4mods}}
\begin{figure}[h!]
\centering
\includegraphics[width=0.7\textwidth]{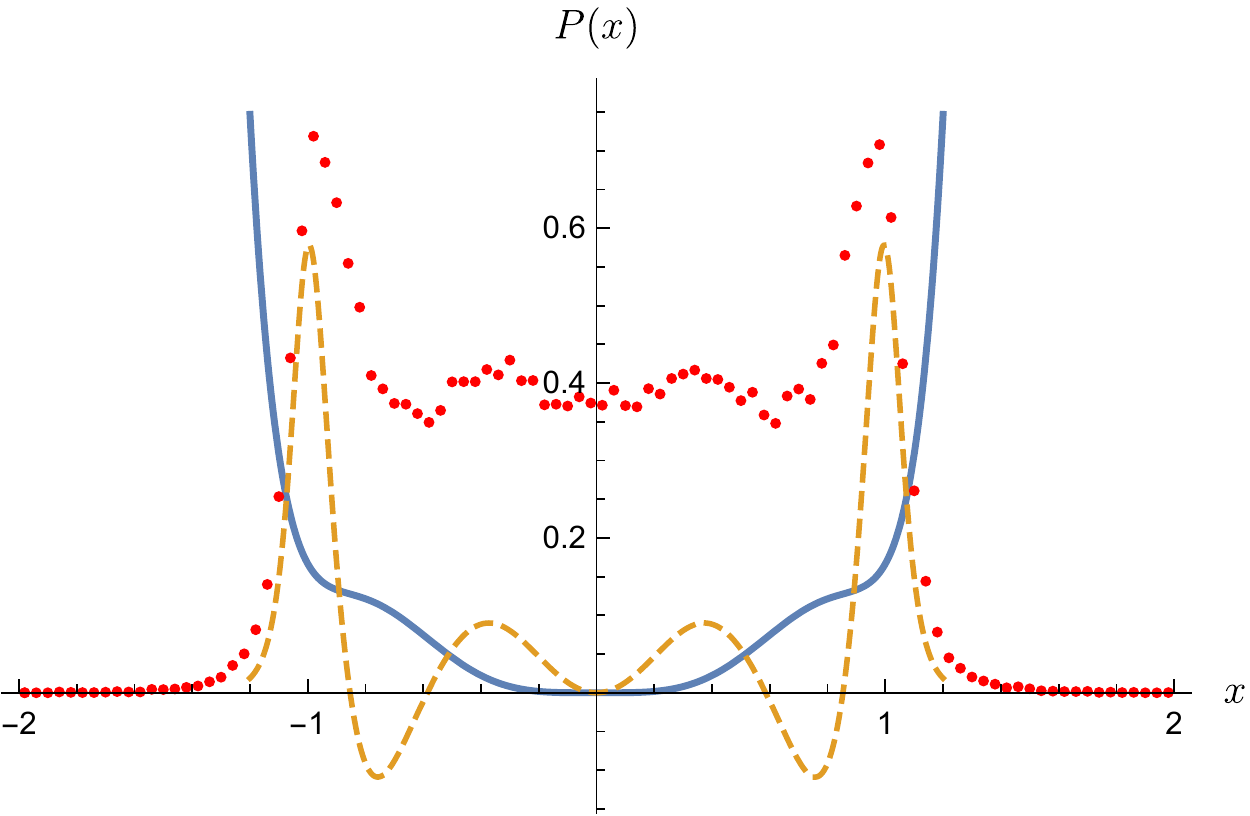}
\caption{The stationary state for the potential \eqref{eq:V2} subject to the $\alpha$-stable driving with $\alpha$=1 (Cauchy noise) and $\sigma=1$ (dots),  potential profile (blue solid line) and the potential curvature (orange dashed line).
Simulation parameters: $N=9.6 \times 10^4$, $\Delta t =10^{-6}$, $\tmax=100$.}
\label{fig:4mods}
\end{figure}

As a sample potential with the single minimum located at  $x=0$ and four maxima of the curvature we use
\begin{equation}
    V(x)=\frac{7}{6}x^4 - 2x^6 +x^8.
    \label{eq:V2}
\end{equation}
Since the potential (\ref{eq:V2}) has four maxima of the curvature, we are expecting that a stationary state will have four modal values.
Indeed, in Fig.~\ref{fig:4mods} four maxima of the stationary density located near points of the maximal curvature are visible.
Maxima located at $x \approx \pm 1$ are significantly higher than maxima at $x \approx \pm 0.4$, because the curvature at these points is substantially smaller. 
In contrast to the potential considered in the previous subsection, the probability density has minimum at $x=0$ due to minimum of the potential curvature.
Alternatively, shape of the stationary density around $x=0$ can be explained by the analysis of the potential.
The potential given by Eq.~\eqref{eq:V2}, for small $x$, behaves like $x^4$ while the potential given by Eq.~\eqref{eq:V1} like $x^2$ resulting in the maximum of stationary density in the former case and minimum of stationary denity in the current case.

%%%%%%%%%%%%%%%%%%%%%%%%%%%%
%
% fig
%

%%%%%%%%%%%%%%%%%%%%%%%%%%%%%%%%%%%%%%%%%%%%%%%%%%%%%
\pagebreak
\subsubsection*{Five mods\label{sec:5mods}}

\begin{figure}[h!]
\centering
\includegraphics[width=0.7\textwidth]{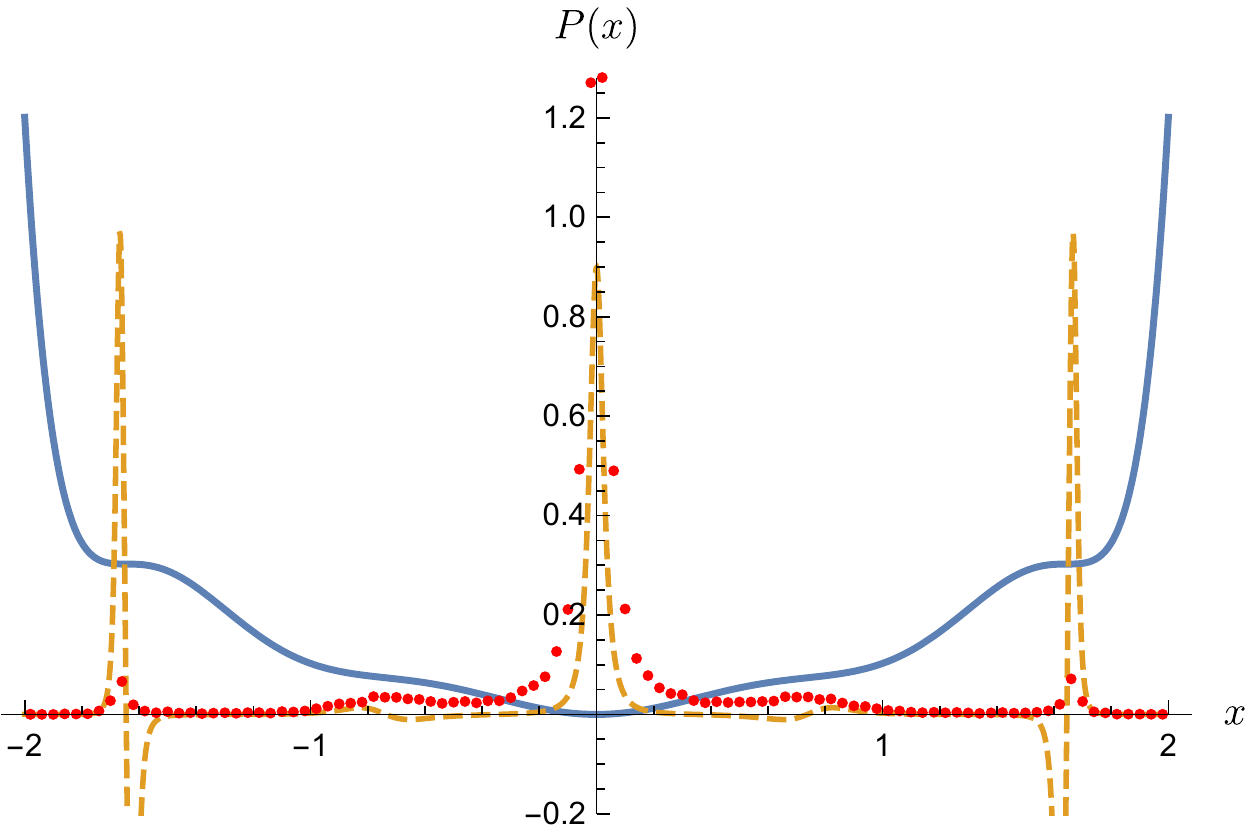}
\caption{The stationary state for the potential \eqref{eq:V3} subject to the $\alpha$-stable driving with $\alpha$=1 (Cauchy noise) and $\sigma=1$ (dots),  potential profile (blue solid line) and the potential curvature (orange dashed line).
Simulation parameters: $N=9.6 \times 10^4$, $\Delta t =10^{-6}$, $\tmax=100$.}
\label{fig:5mods}
\end{figure}

\begin{figure}[h!]
\centering
\includegraphics[width=0.7\textwidth]{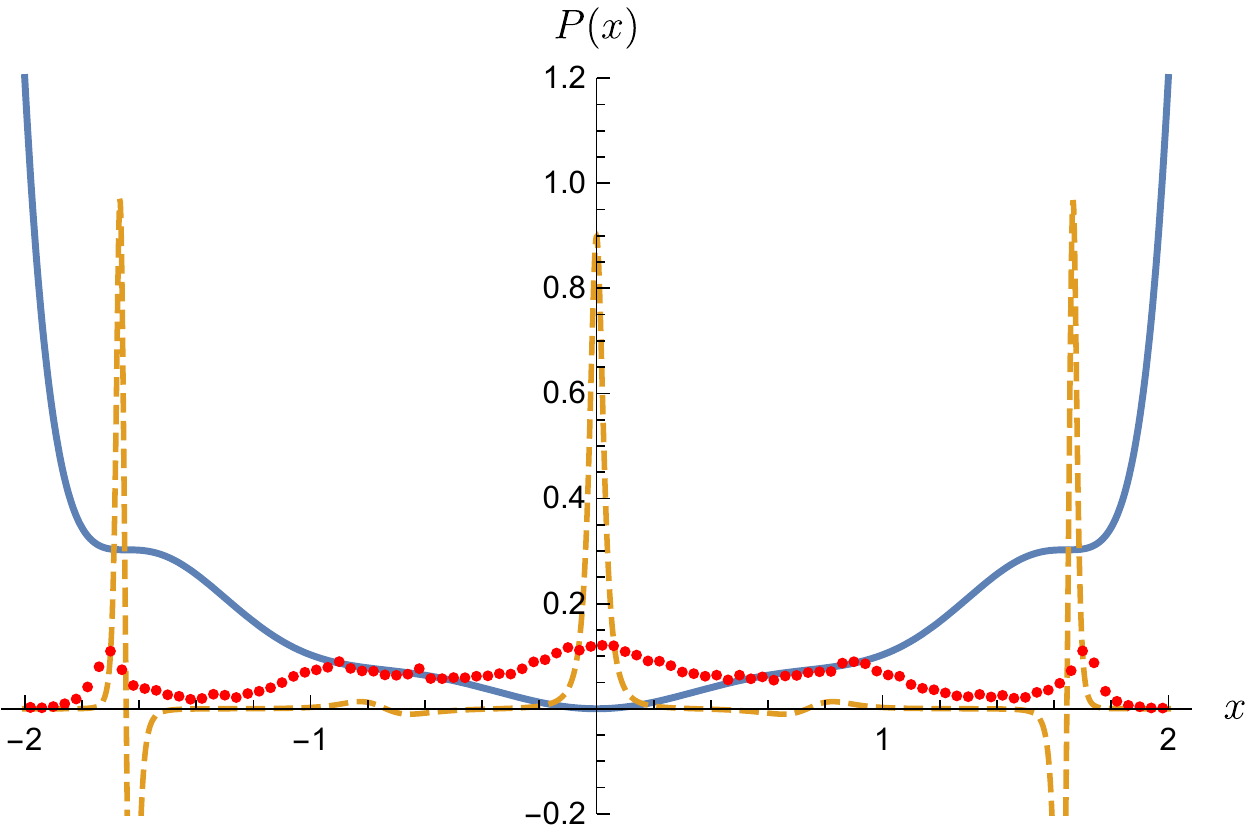}
\caption{The same as in \figref{fig:5mods} for $\sigma=10$.}
\label{fig:5modsD10}
\end{figure}

In order to produce five modal values in the stationary state we use a potential
\begin{eqnarray}
      V(x) & = & 13.4789 x^2-24.8828 x^4+22.6289 x^6 \nonumber \\
      && -8.125 x^8+x^{10}
    \label{eq:V3}
\end{eqnarray}
having five maxima of the curvature.
As it is confirmed by Fig.~\ref{fig:5mods}, the stationary state corresponding to the potential (\ref{eq:V3}) has five modal values.
Additional Fig.~\ref{fig:5modsD10} examines in more details the sensitivity of the stationary state to the scale parameter $\sigma$, which is ten times larger than in Fig.~\ref{fig:5mods}.
Increase in the scale parameter decreases the height of the central maximum and spreads outer peaks.
The decrease of the central peak in the stationary state is related to the central part of distribution of random pulses.
More precisely, for larger values of the scale parameter $\sigma$, peaks of $\alpha$-stable densities become lower and wider transferring effectively a part of the probability mass to tails of distributions.
This in turn, can spread outer peaks of stationary densities and increase their height as it can be deducted from comparison of Figs.~\ref{fig:5mods} and~\ref{fig:5modsD10}.

%%%%%%%%%%%%%%%%%%%%%%%%%%%%
%
% fig
%

%%%%%%%%%%%%%%%%%%%%%%%%%%%%
%
% fig
%

\begin{figure}[h!]
\centering
\includegraphics[width=0.7\textwidth]{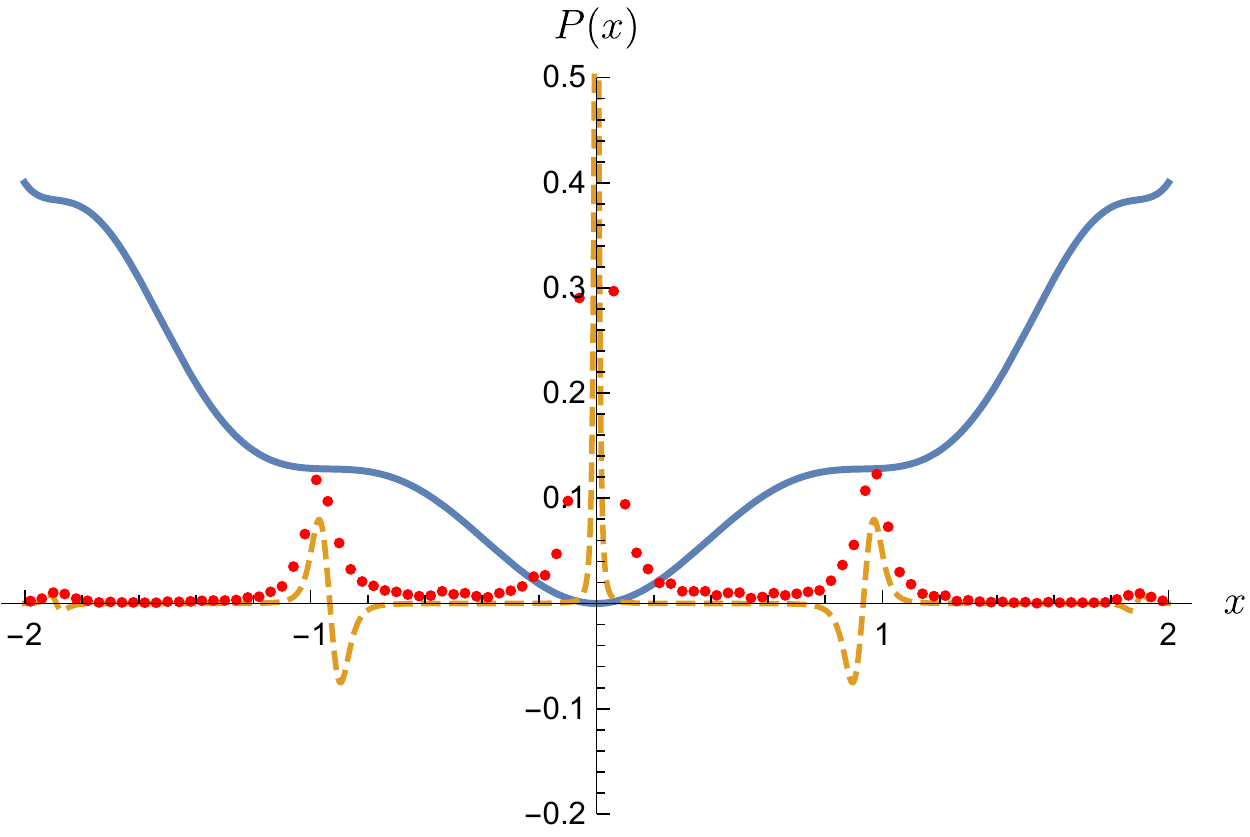}
\caption{The stationary state for the potential \eqref{eq:V4} subject to the $\alpha$-stable driving with $\alpha$=1 (Cauchy noise) and $\sigma=1$ (dots),  potential profile (blue solid line) and the potential curvature (orange dashed line).
Simulation parameters: $N=9.6 \times 10^4$, $\Delta t =10^{-6}$, $\tmax=100$.}
\label{fig:5modsII}
\end{figure}

\begin{figure}[h!]
\centering
\includegraphics[width=0.7\textwidth]{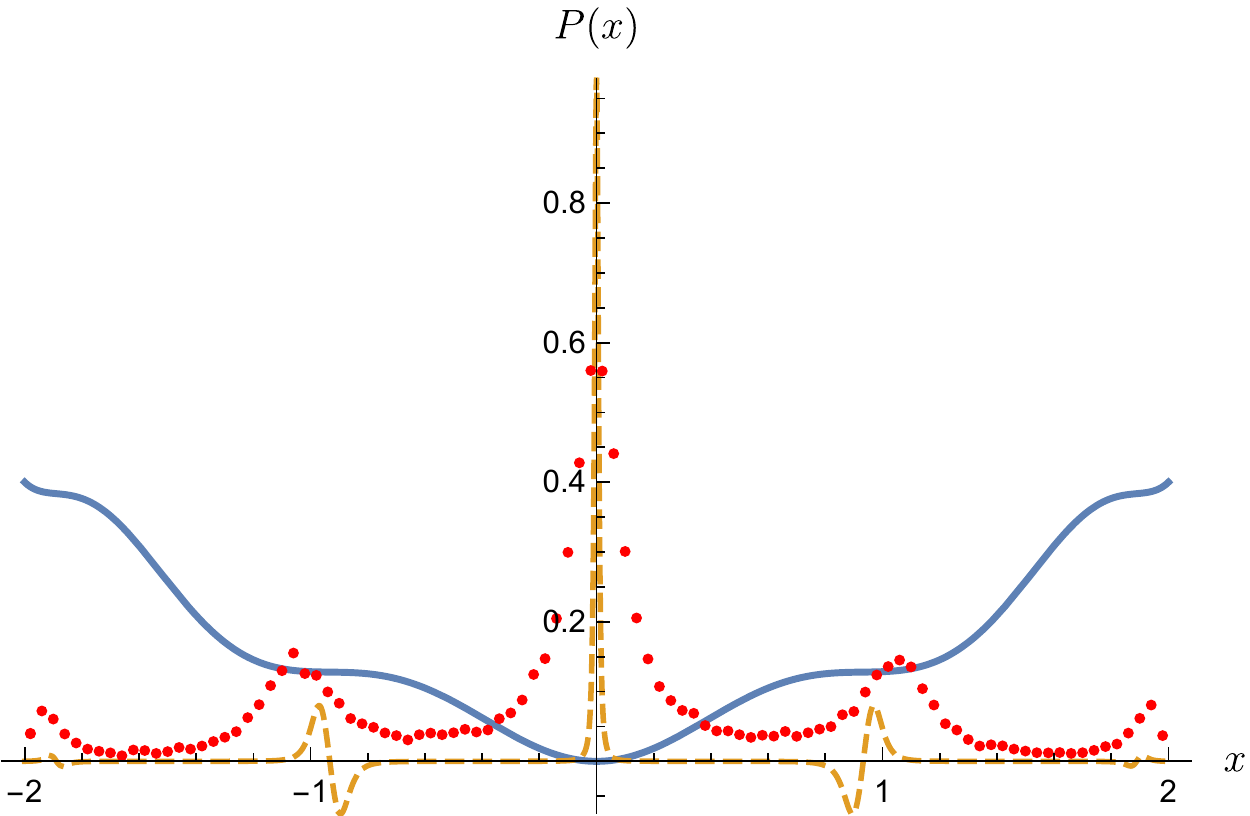}
\caption{The same as in \figref{fig:5modsII} for $\sigma=10$.}
\label{fig:5modsIID10}
\end{figure}

Next, considered sample potential which allow for the emergence of a five-modal stationary state can be
\begin{eqnarray}
  V(x) & = &  49.0625 x^2-69.5313 x^4+43.4414 x^6 \nonumber \\
   & & -11.125 x^8+x^{10}.
  \label{eq:V4}
\end{eqnarray}
Stationary states corresponding to the potential~(\ref{eq:V4}) are presented in Figs.~\ref{fig:5modsII} and  \ref{fig:5modsIID10}.
As in the previous case, see Eq.~(\ref{eq:V3}), the increase in the scale parameter $\sigma$ decreases the height of the central peak and makes outer peaks more pronounced.
Therefore, for large $\sigma$ in the stationary state there are five well visible peaks, see Fig.~\ref{fig:5modsIID10}.

%%%%%%%%%%%%%%%%%%%%%%%%%%%%
%
% fig
%

%%%%%%%%%%%%%%%%%%%%%%%%%%%%%%%%%%%%%%%%%%%%%%%%%%%%%
%%%%%%%%%%%%%%%%%%%%%%%%%%%%%%%%%%%%%%%%%%%%%%%%%%%%%
\subsection{``Glued'' potentials\label{sec:non-dif}}

\begin{figure}[h!]
\centering
\includegraphics[width=0.7\textwidth]{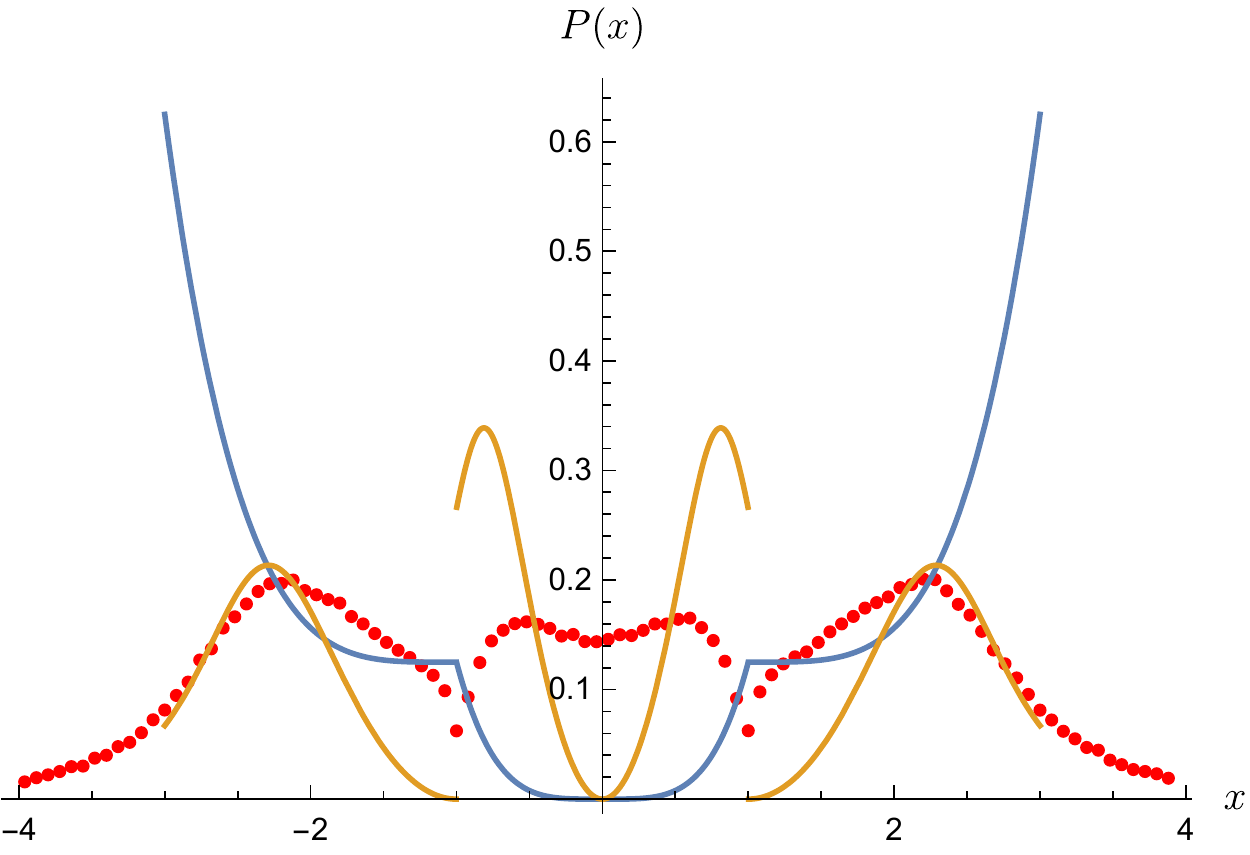}
\caption{
The stationary state for the potential given by \eqref{eq:step} subject to the $\alpha$-stable driving with $\alpha$=1 (Cauchy noise) and $\sigma=1$  (dots), the potential profile (blue solid line) and the potential curvature (orange dashed line). Simulation parameters: $N=10^6$, $\Delta t =10^{-3}$, $\tmax=100$.}
\label{fig:5modsIV}
\end{figure}

Stationary states with more than two modal values can be also produced in continuous non-differentiable ``glued'' potentials.
For instance, let us consider the potential given by
\begin{equation}
    V(x)=
    \left\{
    \begin{array}{lcl} 
    \frac{(x+1)^4}{16}+\frac{1}{4} & \mbox{for} & x < 1 \\
     \frac{x^4}{4} & \mbox{for} & |x|\leqslant  1 \\
     \frac{(x-1)^4}{16}+\frac{1}{4} & \mbox{for} & x > 1  
    \end{array}
    \right.
    .
    \label{eq:step}
\end{equation}
The stationary state, along with the potential profile and the potential curvature is depicted in Fig.~\ref{fig:5modsIV}.
The potential~(\ref{eq:step}) consist of tailored $x^4$ parts.
For small $x$, the stationary state is determined by the part of the potential located at $|x|<1$. 
This part of the potential is quartic, thus it produces two maxima.
Two further maxima, at larger absolute values of arguments, are produced by outer parts of the potential well.
On the one hand, one can see the stationary state as an outcome of competition between two parts (inner and outer) of the potential, which are responsible for the emergence of modal values.
On the other hand, it is possible to provide alternative explanation based on competition between deterministic and random forces.
The inner part of the potential produces two maxima in an analogous way like the $x^4/4$ potential, see Fig.~\ref{fig:n4}.
If a particle, due to random force, escapes to a distant point it starts to slide down towards $x=0$. 
The time scale associated with the deterministic sliding is infinite making probability mass to concentrate on almost horizontal part of the potential close to $|x|=1$ giving rise to two outer maxima of the stationary state, see Fig.~\ref{fig:5modsIV}.
For a potential given by Eq.~(\ref{eq:step}), replacement of the quartic part at $|x|<1$ with the parabolic potential destroys inner maxima of the stationary state.
The inner part of the stationary density is uni-modal, which is consistent with results for the parabolic potential.
Finally, also the phenomenological interpretation based on the potential curvature works.
Maxima of the stationary states are located close to maxima of the potential curvature.

\begin{figure}[h!]
\centering
\includegraphics[width=0.7\textwidth]{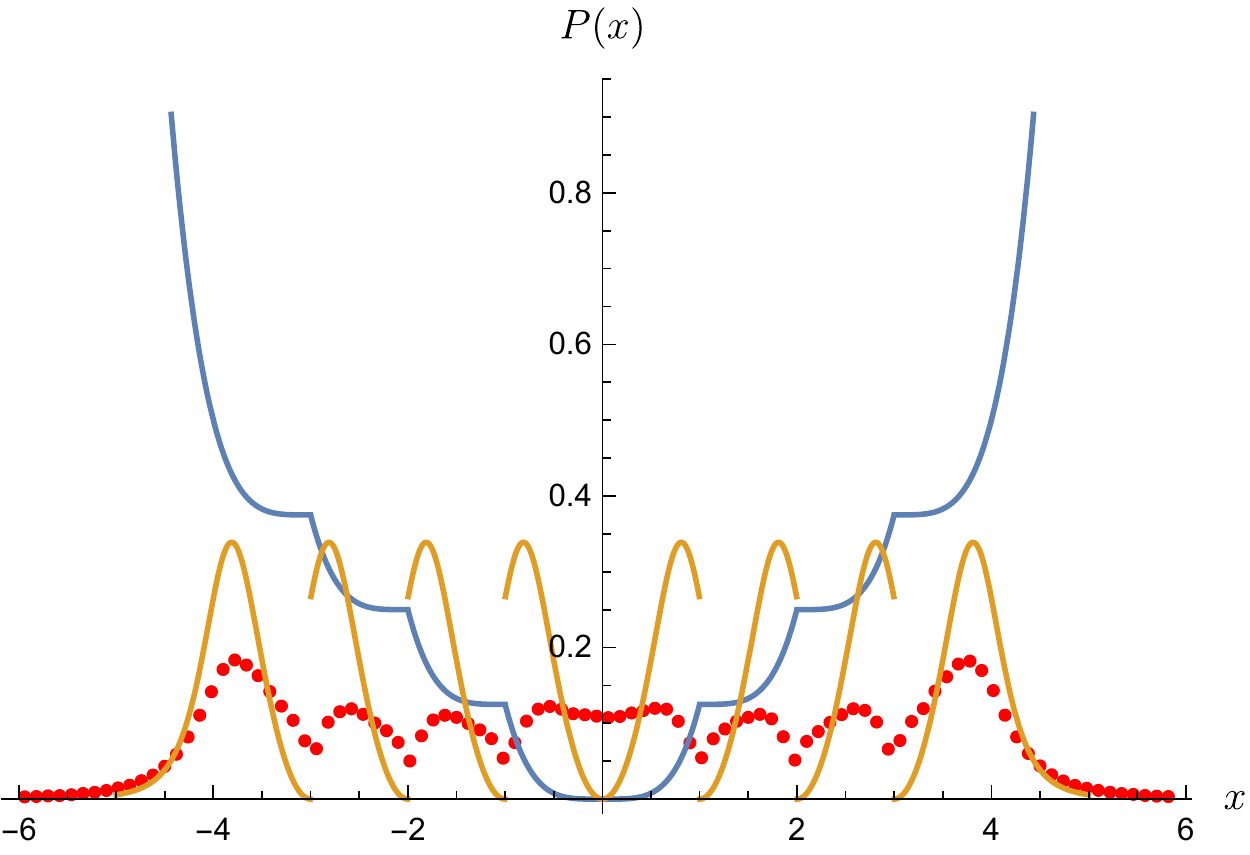}
\caption{The same as in \figref{fig:5modsIV} for the potential given by Eq.~\eqref{eq:step8}.}
\label{fig:8mods}
\end{figure}

The mechanism responsible for emergence of maxima of the stationary state in Fig.~\ref{fig:5modsIV} can be further extended.
For example, the following potential 
\begin{equation}
    V(x)= \left\{
    \begin{array}{lcl} 
  \frac{1}{4} (x+3)^4+\frac{3}{4}& \mbox{for} & x<-3\\
 \frac{1}{4} (x+2)^4+\frac{1}{2} & \mbox{for} & -3 \leqslant x < -2\\
 \frac{1}{4} (x+1)^4+\frac{1}{4}& \mbox{for}  &  -2 \leqslant x < -1   \\
 \frac{x^4}{4} & \mbox{for} & \left| x\right| \leqslant 1 \\
  \frac{1}{4} (x-1)^4+\frac{1}{4} & \mbox{for} & 1<x\leqslant 2 \\
 \frac{1}{4} (x-2)^4+\frac{1}{2} & \mbox{for} & 2<x\leqslant 3 \\
   \frac{1}{4} (x-3)^4+\frac{3}{4} & \mbox{for} & x>3 \\
\end{array}
\right.
    \label{eq:step8}
\end{equation} 
results in emergence of eight-modal stationary state, see Fig.~\ref{fig:8mods}.
The procedure of tailoring potentials, see Eqs.~(\ref{eq:step}) and (\ref{eq:step8}), can be further continued.
Please note, however, that outer maxima are the strongest because they aggregate particles sliding down from the whole outer parts of the potential.

%%%%%%%%%%%%%%%%%%%%%%%%%%%%
%
% fig
%

%%%%%%%%%%%%%%%%%%%%%%%%%%%%
%
% fig
%

%%%%%%%%%%%%%%%%%%%%%%%%%%%%%%%%%%%%%%%%%%%%%%%%%%%%%%%%%%%%%%%%%%%%%
%%%%%%%%%%%%%%%%%%%%%%%%%%%%%%%%%%%%%%%%%%%%%%%%%%%%%%%%%%%%%%%%%%%%%%%
\section{Summary and Conclusion\label{sec:summary}}

It is well known that stationary states in systems driven by L\'evy noise can display intriguing properties \cite{chechkin2002,chechkin2003,chechkin2004,chechkin2006,chechkin2008introduction}.
Almost twenty years ago it was proved that stationary states of non-harmonic L\'evy, e.g. quartic, oscillators can be bi-modal \cite{chechkin2002}.
Nevertheless, so far, it has not been verified if stationary states in symmetric single-well potential can be characterized by more than two modal values.
The current manuscript provides conclusive and positive answer to this problem.

We have grounded our considerations on a condition  of existence of modal values which attribute maxima of probability density to maxima in the potential curvature \cite{chechkin2004}.
In the next step this conjecture has been used to construct symmetric single-well potentials resulting in desired multimodality.
Finally, we have numerically investigated properties of stationary states proving that stationary states indeed have anticipated multimodality.
\bd{This step is necessary because the curvature condition might be not sufficient to acquire anticipated multi-modality. 
Therefore, the final test and fine-tuning of potential parameters need to be performed manually.}
All considered potential assured existence of stationary states, because every potential for large $|x|$ is steeper than $x^4$ which is well above the minimal steepness ensuring existence of stationary states.

Peaks in the stationary state which are located close to the origin are determined by the behaviour of the potential at small $x$.
Therefore, if the dominating part of the potential at $x \approx 0$ is steeper than parabolic the stationary state has a minimum at the origin what is especially well visible for the  quartic potential.
In contrast, for potentials less steep than parabolic, the stationary state has a global maximum at the origin.
In general, modal values of stationary states are located in the vicinity of maxima of the potential curvature.
The mechanism of emergence of multimodal stationary states is better visible for ``glued'' potentials than for continuous differentiable potentials.

Within simulations, we have focused on the Cauchy noise which is a special example of the L\'evy noise with the stability index $\alpha=1$.
However, similar considerations can be performed for other allowed values of the stability index.
In the limiting case of $\alpha=2$, the $\alpha$-stable noise becomes the Gaussian white noise.
Therefore, stationary states become of the Boltzmann-Gibbs type and for single-well potentials they are single-modal.

\bd{
Obtained findings indicate that the stationary state in a single-well potential can be of non-trivial, multimodal type. Therefore, it might be important to asses the role of multi-modal stationary states in noise-induced effects. This problem seems to be especially relevant in the context of averaging over equilibrium initial conditions, sensitivity to initial conditions or the problem of transition over the potential barrier.
For the Kramers problem, the issue of placing the boundary which discriminate  states \cite{dybiec2007b,dybiec2007d} might be further complicated due to multi-modality of stationary states.
}

%%%%%%%%%%%%%%%%%%%%%%%%%%%%%%%%%%%%%%%%%%%%%%%%%%%%%%%%%%%%%%%%%%%%%%%%%%%%%%%%%%%%%%%%%
%%
%% ACKNOWLEDGMENTS
%%
% \newpage
\ack
This project was supported by the National Science Center grant (2014/13/B/ST2/02014). 
Computer simulations have been performed at the Academic Computer Center Cyfronet, AGH University of Science and Technology (Krak\'ow, Poland) under CPU grant ``DynStoch''.
Fruitful suggestions from Aleksei Chechkin are greatly acknowledged.

% \section*{References}
% \bibliographystyle{iopart-num}
% \bibliography{core-bibliography}

\def\url#1{}
\providecommand{\newblock}{}

\end{document}